\documentclass[dvips]{article}

\usepackage{icrc2011}

\usepackage{url}
\usepackage{multirow}

\title{Development of an ASIC for Dual Mirror Telescopes of the Cherenkov Telescope Array}

\newcommand{\etal}{\MakeLowercase{\textit{et al. }}} 
\shorttitle{J.~Vandenbroucke \etal Development of TARGET}

\authors{Justin Vandenbroucke$^{1}$, Keith Bechtol$^{1}$, Stefan Funk$^{1}$, Akira Okumura$^{2, 1}$, Hiro Tajima$^{3}$,\\ Gary~Varner$^{4}$, for the CTA consortium}
\afiliations{$^1$W. W. Hansen Experimental Physics Laboratory, Kavli Institute for Particle Astrophysics and Cosmology, Department of Physics and SLAC National Accelerator Laboratory, Stanford University, Stanford, CA 94305, USA
\\
$^2$Institute of Space and Astronautical Science, JAXA, Sagamihara, Kanagawa 252-5210, Japan\\
$^3$Solar-Terrestrial Environment Laboratory, Nagoya University, Nagoya 464-8601, Japan\\
$^4$Department of Physics and Astronomy, University of Hawaii, Honolulu HI 96822, USA}

\email{justinv@stanford.edu (Justin Vandenbroucke)}

\abstract{We have developed an application-specific integrated circui (ASIC) for photomultipler tube (PMT) waveform digitization which is well-suited for the Schwarzschild-Couder optical system under development for the Cherenkov Telescope Array (CTA) project.  The key feature of the ``TARGET'' ASIC is the ability to read 16 channels in parallel at a sampling speed of 1 GSa/s or faster.  In combination with a focal plane instrumented with 64-channel multi-anode PMTs (MAPMTs), TARGET digitizers will enable CTA to achieve a wider field of view than the current Cherenkov telescopes and significantly reduce the cost per channel of the camera and readout electronics. We have also developed a prototype camera module, consisting of 4 TARGET ASICs and a 64-channel MAPMT.  We report results from performance testing of the camera module and of the TARGET ASIC itself.}
\keywords{ASIC, Instrumentation, CTA, MAPMT, TARGET}

\begin{document}
\maketitle

\section{Introduction}
\quad The Cherenkov Telescope Array (CTA) experiment is a next-generation verhy-high-energy gamma-ray observatory featureing an array of imaging atmospheric Cherenkov telescopes (IACTs) that will be an order of magnitude more sensitive than the current generation of instruments \cite{The-CTA-Consortium:2010:Design-Concepts-for-the-Cherenkov-Telesc}. The energy band covered by CTA will range from a few tens of GeV to beyond 100~TeV. To achieve the highest gamma-ray sensitivity ever with this wide energy coverage, CTA will be an array of $\sim100$ telescopes consisting of a mix of a few different telescope designs.

\quad One candidate of telescope designs is Schwarzschild-Couder midium-size telescope (SC-MST) which is being developed to realize a wide field of view (FOV) ($\sim8^\circ$ in diameter) and high angular resolution ($<\sim0.1^\circ$) at the same time by using dual mirrors in the optical system \cite{Vassiliev:2007:Wide-field-aplanatic-two-mirro}. The focal-plane camera of the SC optical system consists of an array of 64-channel multi-anode photomultiplier tubes (MAPMTs), because the f-ratio of the SC optics is a few times larger than those of normal IACTs, and thus the pixel size of the camera is required to be smaller than regular PMTs with diamters of $\sim25\ \mathrm{mm}$.

\quad In order to read Cherenkov signals from an MAPMT array, a compact and modular readout system running at a sampling speed of $>\sim1$ GSa/s (giga-samples per second) is required. In addition, the cost per channel of the readout system is required to be as low as possible because a large number of telescopes are to be built.

\section{TARGET}

\quad We have developed an application-specific integrated circuit (ASIC) which was designed to match the requirements of the SC-MST. The first generation of this ASIC, TeV Array Readout with GSa/s sampling and Event Trigger (TARGET~1), has self-trigger functionality, 16-channel parallel input, and a 4096-sample buffer for each channel \cite{Bechtol:2011:TARGET:-A-multi-channel-digitizer-chip-f}. The total cost per channel including front-end and back-end electronics is expected to be $\sim\$20$ not including photo detectors.

\begin{figure*}[tb]
  \centering
  \includegraphics[width=6.5in]{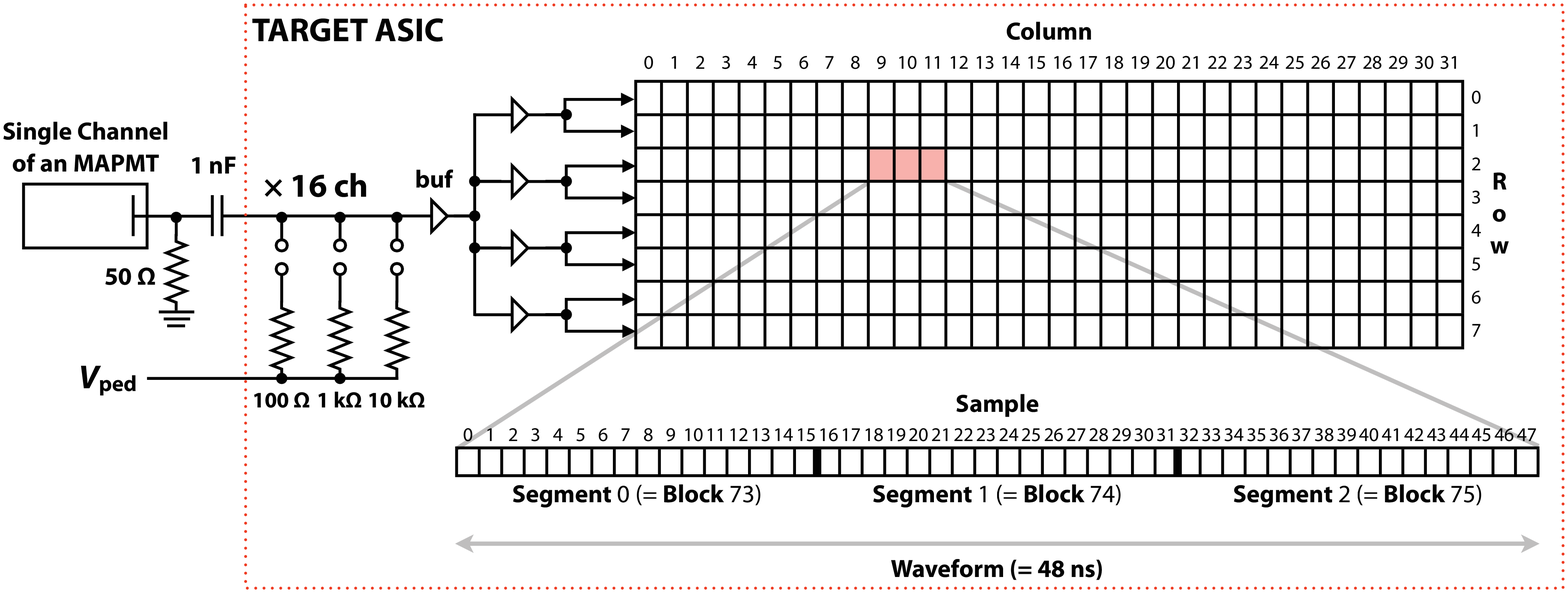}
  \caption{A schematic diagram of the TARGET~1 ASIC. A single AC-coupled channel of an MAPMT is connected to an array of $4096$ capacitors. The 16 channels are digitized in pairs, with the two channels of each pair digitized simultaneously. Three input impedances are selectable via the FPGA firmware. The pedestal level can also be changed by changing an external voltage, $V_{\mathrm{ped}}$.}
  \label{fig_TARGET}
\end{figure*}

\quad Figure~\ref{fig_TARGET} illustrates a schematic diagram of the TARGET~1 ASIC which has an array of $4096$ capacitors divided into 256 blocks aligned in 32 columns by 8 rows, where each block consists of 16 capacitors. The sampling speed of the array can be adjusted between $0.7$~GSa/s and $2.3$~GSa/s by changing an external voltage input, but it is typically driven at $1.0$~GSa/s. Therefore, each capacitor and the total buffer depth correspond to $1$~ns and $4096$~ns, respectively. By reading three blocks upon each trigger, the waveform length becomes $48$~ns, while the length can be changed through a field-programmable gate array (FPGA). We typically set the waveform length to $48$ or $64$~ns in testing.

\quad The storage capacitor voltage is digitized by Wilkinson-type analog-to-digital converters (ADC) equipped in TARGET~1. The voltage is measured by a gray-code counter which starts at the beginning of the Wilkinson ramp voltage, and stops when the ramp voltages equals the capacitor voltage. In the TARGET~1 evaluation board and camera module prototype to be explained in Section~\ref{sec_eval_camera}, the ADC resolution is 10~bits and 9~bits, respectively.

\quad Table~2 is a summary of the specifications of TARGET~1. The same parameters for the 2nd version of TARGET (TARGET~2) are also listed. TARGET~2 chips have been designed and fabricated, and will be tested in 2011.

\begin{table*}[tb]
  \caption{Performance parameters of TARGET~1 and TARGET~2 \cite{Bechtol:2011:TARGET:-A-multi-channel-digitizer-chip-f}. The sampling frequency, bandwidth, and cross talk of TARGET~1 are based on actual laboratory measurements, while those of TARGET~2 are simulated.}
  \label{table_spec}
  \begin{center}
    \begin{tabular}{l|c|c}
      \hline
      Parameter & TARGET 1 & TARGET 2\\
      \hline\hline
      Channels & $16$ & $16$ \\
      Dynamic range (bits) & $9$ or $10$ & up to $12$ \\
      Sampling frequency (GSa/s) & $0.7-2.3$ & $0.2-1.8$ \\
      3 dB analog bandwidth (MHz) & $150$ & $>380$ \\
      Cross talk at 3 dB frequency & $<4$\% & $1$\% \\
      Buffer depth (cells per channel) & $4,096$ & $16,384$ \\
      Wilkinson ADC counter speed (MHz) & $445$ & $700$ \\
      Samples per digitization (block size) & $16$ & $32$ \\
      Digitization time per block ($\mu$s) & $1$ (9-bit) or 2 (10-bit) & 0.7 (9-bit) or 1.5 (10-bit) \\
      Number of cells digitized simultaneously & $16$ cells $\times$ $2$ channels & $32$ cells $\times$ $16$ channels \\
      Clock speed for serial data transfer (Mbps) & -- & $100$ \\
      Channels for simultaneous data transfer & -- & $16$ \\
      Dead time for $48$ samples $\times$ $16$ ch ($\mu$s) & \multirow{2}{*}{$24+0$ (9-bit) or $48+0$ (10-bit)} & \multirow{2}{*}{$1.5+7.2$ (9-bit) or $2.9+7.2$ (10-bit)} \\
      (Digitization time $+$ readout time )& & \\
      Trigger outputs & $1$ (OR of $16$ channels) & $4$ (each is analog sum of $4$ channels) \\
      \hline
    \end{tabular}
  \end{center}
\end{table*}

\section{The TARGET~1 Camera Module Prototype}
\label{sec_eval_camera}

\begin{figure}[tb]
  \vspace{5mm}
  \centering
  \includegraphics[width=3.2in]{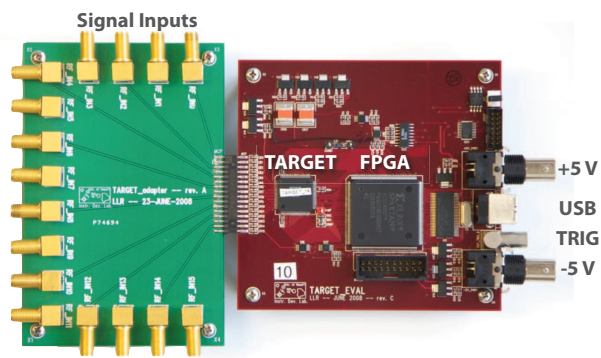}
  \caption{A photo of the TARGET~1 evaluation board (right) and 16-channel input board (left). Basic characteristics of TARGET~1 were measured with the evaluation board. Its main components are an ASIC, an FPGA, $\pm5$~V DC input, and a USB interface to be connected to a computer.}
  \label{fig_EvalBoard}
\end{figure}

\begin{figure}[tb]
  \vspace{5mm}
  \centering
  \includegraphics[width=3.2in]{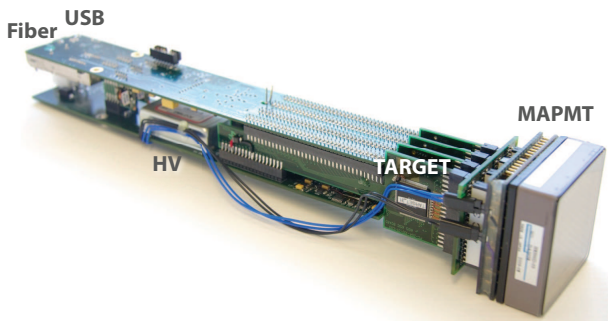}
  \caption{A photo of the TARGET~1 camera module. A TARGET~1 chip is located on each of four vertical boards connected to a 64-ch MAPMT. An FPGA is located on the backside of the top board.}
  \label{fig_CameraModule}
\end{figure}

\quad We have also developed the TARGET~1 evaluation board and the TARGET~1 camera module prototype as shown in Figure~\ref{fig_EvalBoard} and~\ref{fig_CameraModule}. Since the evaluation board was fabricated to study the basic characteristics of a TARGET~1 chip, it consists of a minimum set of components to operate a single ASIC. The camera module was designed to validate the concept of a combination of a 64-channel MAPMT and 4 TARGET~1 chips (16 channels by 4 ASICs). $\sim200$ camera modules will be installed on the focal-plane camera of a single SC telescope in the future, where each $\sim36$ modules will be controlled separately by a backplane board.

\quad The camera module prototype has an MAPMT\footnote{Hamamatsu Photonics H8500D-03 is currently used. It can be replaced with an array of multi-pixel photon counters (MPPCs) in the future.}, a high-voltage (HV) power unit\footnote{Negative HV for an MAPMT and positive for an MPPC array}, a universal serial bus (USB) interface, a fiber optic interface, 4 separate ASIC boards, and an FPGA. The fiber optic interface enables us to acquire 64-channels waveforms from two camera modules at a rate of $>3.3$~kHz in the current design, while the speed of the USB interface is only $\sim40$~Hz. The USB interface is used in initial testing at low trigger rate. When multiple camera modules are operated by a subfield board at higher trigger rate, the fiber interface is used instead.

\section{Performance Tests}

\begin{figure}[tb]
  \vspace{5mm}
  \centering
  \includegraphics[width=3.2in]{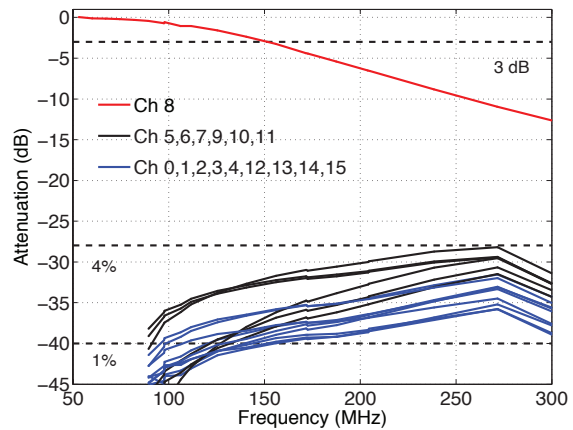}
  \caption{Analog bandwidth and cross talk of TARGET~1. A sinusoidal signal was input to only channel 8, and all 16 channels were read. Channel 8 data (red) shows that the attenuation is $-3$~dB at 150~MHz. Cross talk is 4\% at maximum when $\sim250$~MHz signal is given.}
  \label{fig_CrossTalk}
\end{figure}

\quad Many tests have been done using the evaluation board and the camera module, details of which are fully covered elsewhere \cite{Bechtol:2011:TARGET:-A-multi-channel-digitizer-chip-f} (hereafter TARGET~1 paper). Figure~\ref{fig_CrossTalk} is one of the  measurements, showing the analog bandwidth of TARGET~1 against sinusoid input at various frequencies. The measured bandwidth, $-3$~dB at $\sim150$~MHz, is somewhat low, but it is expected to be improved to $>380$~MHz in TARGET~2 (see Table~\ref{table_spec}).

\begin{figure}[tb]
  \vspace{5mm}
  \centering
  \includegraphics[width=3.in]{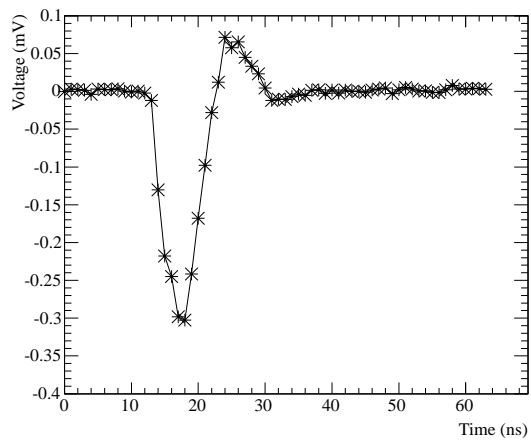}
  \caption{An example waveform of the MAPMT digitized by the camera module prototype. The waveform length is 64~ns (4 blocks) in this example.}
  \label{fig_waveform}
\end{figure}

\begin{figure}[tb]
  \vspace{5mm}
  \centering
  \includegraphics[width=3.in]{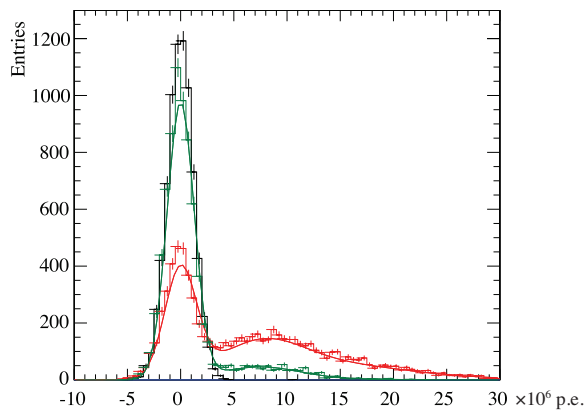}
  \caption{Single photoelectron distribution of the MAPMT obtained by the camera module prototype. Black line shows the pedestal distribution (no signal), green and red show the single photoelectron peak. Darker ND filter was used in the green data.}
  \label{fig_1pe}
\end{figure}

\quad Figures~\ref{fig_waveform} and~\ref{fig_1pe} show an example of waveform and single photoelectron distribution of the MAPMT, respectively, which were digitized using the camera module. The self-trigger functionality, waveform digitization, and data-stream chain from MAPMT to an external data-acquisition computer have been validated.

\quad Since the design of the backplane board is not completed yet, we are testing the fiber interface using another project's board temporarily which uses the same data transfer protocol as the module. In the current configuration, the board is able to receive  waveforms from two camera modules simultaneously. Using the temporary board, we achieved an event rate of $3.3$~kHz when taking in total, $64\times2$ waveforms from two camera modules. The speed can be improved to be faster with a faster optic interface and firmware upgrade.

\section{Conclusion and Prospects}
\quad The TARGET~1 chip and the camera module prototype have been fabricated and their performance has been well evaluated. We found that most performance characteristics meet our requirements for a SC-MST. However, a couple of problems, such as low bandwidth and AC-linearity saturation against high-frequency signals ($>50$~MHz), are known as reported in the TARGET~1 paper. Digitization noise in the camera module, which is not negligible, is also known, but it is not an intrinsic problem of TARGET~1. Those problems are to be removed in TARGET~2 and the second version of the camera module to be developed. The new system will be tested in 2011 and 2012.

\quad In addition to the TARGET development, SC telescopes, subfield boards, back-end electronics, MAPMTs, and full Monte Carlo simulations are also being studied and developed in the CTA collaboration.

\section{Acknowledgments}
\quad We gratefully acknowledge support from the agencies and organisations listed in this page: \url{http://www.cta-observatory.org/?q=node/22}

\end{document}